\documentclass[fleqn,usenatbib]{mnras}

\usepackage[T1]{fontenc}

\DeclareRobustCommand{\VAN}[3]{#2}
\let\VANthebibliography\thebibliography
\def\thebibliography{\DeclareRobustCommand{\VAN}[3]{##3}\VANthebibliography}

\usepackage{graphicx}	
\usepackage{amsmath}	
\usepackage{amssymb}	
\usepackage{natbib}
\usepackage{xcolor}     

\newcommand\tmin{$T_{\rm min}$}
\newcommand\tmax{$T_{\rm max}$}
\newcommand\fpdr{$f_{\rm PDR}$}

\newcommand\tmw{$T_{\rm mw}$}

\newcommand\minput{$M_{\rm input}$}
\newcommand\tfit{$T_{\rm fit}$}
\newcommand\mfit{$M_{\rm fit}$}

\newcommand\bfit{$\beta_{\rm fit}$}

\usepackage{newtxtext,newtxmath}

\title[Effect of T-dependent dust optical properties]{Effects of temperature-dependent optical properties on the determination of interstellar dust masses}

\author[L. Fanciullo et al.]{
Lapo Fanciullo,$^{1,2,3}$\thanks{E-mail: lfanciullo.astro@gmail.com}
Jonathan P. Marshall,$^{3}$
Francisca Kemper,$^{4,5,6}$ 
Peter Scicluna$^{7,8}$
and Sundar Srinivasan$^{9}$
\\
$^{1}$ National Chung Hsing University, 145 Xingda Rd., South Dist., Taichung City 402, Taiwan, R.O.C.\\
$^{2}$ Tamkang University, 151 Yingzhuan Rd., Tamsui Dist., New Taipei City 251301, Taiwan, R.O.C.\\
$^{3}$ Institute of Astronomy and Astrophysics, Academia Sinica, 11F of AS/NTU Astronomy-Mathematics Building, No.1, Sec. 4, Roosevelt Rd, Taipei 106319, Taiwan, R.O.C.\\
$^{4}$ Institut de Ci\`encies de l'Espai (ICE, CSIC), Can Magrans, s/n, E-08193 Cerdanyola del Vall\`es, Barcelona, Spain\\
$^{5}$ ICREA, Pg. Lluís Companys 23, E-08010 Barcelona, Spain\\
$^{6}$ Institut d'Estudis Espacials de Catalunya (IEEC), E-08860 Castelldefels, Barcelona, Spain\\
$^{7}$ Centre for Astrophysics Research, Department of Physics, Astronomy and Mathematics, College Lane Campus, University of Hertfordshire, Hatfield AL10 9AB, UK\\
$^{8}$ Space Science Institute, 4750 Walnut Street, Suite 205, Boulder, CO 80301, USA\\
$^{9}$ Instituto de Radioastronomía y Astrofísica, Universidad Nacional Autónoma de México. Antigua Carretera a Pátzcuaro \#8701, Ex-Hda. San José de la Huerta 58089. Morelia, Michoacán, México
}

\date{Accepted XXX. Received YYY; in original form ZZZ}

\pubyear{2026}

\begin{document}
\label{firstpage}
\pagerange{\pageref{firstpage}--\pageref{lastpage}}
\maketitle

\begin{abstract}
Accurate measurements of interstellar dust mass are key to answering several astrophysical questions. A common method of obtaining the mass is to fit the far-infrared thermal emission of the dust with a modified blackbody model; however, this method is subject to several systematics. In particular, how temperature-dependent dust optical properties affect fit results has received little attention. We provide the first quantification of this effect based on experimental measurements of optical properties from the scientific literature. We created a grid of synthetic observations for variable-opacity dust and fitted it with a modified blackbody model; the difference between the input properties of synthetic observations and the values derived from the fit provides a measure of the bias induced by the temperature dependence.
We find that fixing the value of the opacity power law index $\beta$ introduces a temperature-dependent bias on the fit, while keeping $\beta$ as a free parameter introduces a bias that depends mainly on the wavelength range used. For instance, depending on the properties of the observed object and on the choice of fit procedure, temperature dependence alone can induce an overestimate of 25-60\% in dust masses at high redshift ($z \sim 8$). Our findings highlight the limitations of power laws as opacity models.
\end{abstract}




\section{Introduction}
\label{sec:introduction}

Interstellar dust, the solid component of the interstellar medium (ISM), is an ensemble of mostly sub-micron sized silicate and carbon grains. Dust plays an important role in observational astrophysics since it absorbs starlight in the ultraviolet to near-infrared range and re-emits energy at longer wavelengths, mainly in the far-infrared and submillimetre. At these wavelengths, the thermal emission of interstellar dust, dominates the spectral energy distribution (SED) of the interstellar medium. This makes dust an essential diagnostic tool of the ISM, and in particular a tracer of interstellar gas, in both its atomic and molecular phase \citep[after conversion via a dust-to-gas ratio; e.g., ][]{Sandstrom+13}. 
One specific issue in which interstellar dust abundance plays a key role is the so called \textit{dust budget crisis}. Galaxies at a redshift $z \sim 5-10$ appear dustier than expected given the formation and evolution timescales for dust sources, i.e., evolved stars and supernovae \citep[][]{Rowlands+14, Forgan+17, Choban+25}. In contrast, JWST has uncovered a population of massive, essentially dust-free galaxies at even higher redshift \citep[$z > 10$,][]{Ferrara+25}. Accurate estimates of dust masses are necessary to understand and explain this phenomenon and its implications for the evolution of the early Universe. 

However, estimates of interstellar dust masses suffer from large systematic uncertainties, largely stemming from fit degeneracies and from our incomplete understanding of dust optical properties and temperature distribution. One common way of obtaining dust mass and temperature is to fit the SED with a single-temperature modified blackbody: 
\begin{equation}
\label{eq_mbb}
    F_\nu (\lambda) = \frac{1+z}{D_L^2} \ M_{\rm dust} \ \kappa(\lambda) \ B_{\rm \nu}(T_{\rm dust}) 
\end{equation}
which is valid in the optically thin case for a dust mass $M_{\rm dust}$ at a temperature $T$ and redshift $z$. The distance factor $(1+z)/D_L^2$, where $D_L$ is the redshift-dependent luminosity distance \citep{Peacock_99}, reduces to $1/D^2$ in the local Universe.
The dust opacity $\kappa(\lambda)$ is often approximated as a power law:
\begin{equation}
\label{eq_kappa}
     \kappa(\lambda) = \kappa_0 \left(\frac{\lambda}{\lambda_0} \right)^{-\beta} 
\end{equation}
where the value $\kappa_0$ is the opacity at a reference wavelength $\lambda_0$, and the value of $\beta$ is typically between 1.5 and 2. In this type of fit, $M_{\rm dust}$, $T_{\rm dust}$ and $\beta$ are free parameters. While more sophisticated models exist \citep[e.g., ][]{DL07,Compiegne+11,Jones+17}, the one we described is widely used for large surveys, especially at high redshift where observational data is limited and $\beta$ -- and sometimes $T_{\rm dust}$ -- are kept fixed \citep[e.g., ][]{Witstok+23, Eales&Ward_24}.

Despite its wide usage, the modified blackbody model has some limitations, due to its simplified nature. A known source of bias is that it assumes a single temperature: in real targets, the existence of multiple dust temperature components on the line of sight or within the beam tends to widen the SED, and a modified blackbody fit results in a lower \bfit\ and \mfit, as well as a higher \tfit, than ideal \citep[][]{Shetty+09b, Sommovigo+Algera_25}.
Another issue lies in the systematic uncertainties on far-infrared and submillimetre dust opacity, for which equation~(\ref{eq_kappa}) only provides a first-order approximation. Laboratory studies of experimental dust analogues, such as silicates and amorphous carbon, reveal a more complex picture \citep[e.g.,][]{Agladze+94, Mennella+98, Boudet+05, Demyk+22}. Firstly, experimental measurements typically result in significantly higher opacities than those used in standard modified blackbody models. Since $M_{\rm dust}$ and $\kappa_0$ are degenerate, as shown in equations~(\ref{eq_mbb}) and (\ref{eq_kappa}), choosing an opacity that is on average lower than the real one would result in overestimating \mfit. Secondly, dust opacity is not always a simple power law. Forcing a power law on $\kappa(\lambda)$ in the fit could therefore introduce a bias in \tfit\ and \mfit. Finally, the optical properties of amorphous materials, which constitute the majority of interstellar dust \citep[e.g.,][]{Kemper+04}, depend on temperature: a material's opacity tends to increase as its temperature increases. A choice of opacity that is appropriate for cold dust will therefore not be appropriate for warm dust, and vice versa.

The possible systematics stemming from the $M_{\rm dust} - \kappa_0$ degeneracy have been noted before \citep[e.g.,][]{Gordon+14} and have been quantitatively discussed in a previous paper \citep[][hereafter paper I]{Fanciullo+20}. On the other hand, the effects of temperature-dependent opacity and of the departure of $\kappa(\lambda)$ from a pure power law have not yet been accurately studied. While \citet{Dale+Helou_02} and \citet{daCunha+08} have used different $\beta$ values for dust at different temperatures, to our knowledge no systematic study exists that uses the available experimental measurements of temperature-dependent opacity.

The aim of this paper is to determine how the temperature dependence of dust opacity and its non-power law nature affect SED fit results, and whether these effects compound with observational limitations. In the present paper, we do not attempt to obtain an absolute calibration of the opacity, which was the object of Paper~I; rather, we want to understand the effect of opacity variations. To do so, we produced synthetic observations using an experimentally-derived dust opacity, then fit it with a standard modified blackbody model, and compared the fit results with the synthetic observation parameters. Throughout the paper, we make the assumption that the optical properties of materials studied in the laboratory are representative of the likely properties of real dust.

\section{Methodology}
\label{sec:methodology}

Our procedure for estimating the bias due to suboptimal opacity models follows the one adopted in Paper~I. First, we created synthetic multi-band, far-infrared/submillimetre photometry for galaxies using experimental values for dust opacity. We then fit the synthetic observations with a single-temperature modified blackbody model, with an opacity in the form introduced in equation~\ref{eq_kappa}. The difference between the model parameters used to create the synthetic observations (dust mass, temperature distribution) and the values recovered by the fit give us an indication of the likely bias in fitting astronomical observations. The scripts used in the procedure are publicly available, as detailed in the Data Availability section.

The main differences between the procedure in Paper~I and in this paper, aside from adopting a temperature-dependent dust opacity, are the following: whereas in Paper~I we calculated synthetic photometry for single-temperature and two-temperature dust models, here we used a more realistic temperature distribution (see Section~\ref{sec:methodology:t_dist}). Also, while in paper~I we adopted the dust opacity from \cite{James+02} as representative of the scientific literature (what could be termed a ``legacy dust model''), here we used a power law fit to the experimental opacity at $T=20$ K. Using this as our fiducial dust opacity eliminates by construction the bias due to an incorrect value of $\kappa_0$, which was the focus of Paper~I. Any remaining bias in the fit results is therefore caused mainly by temperature-dependent variations in the opacity $\kappa(\lambda)$, and by the limitations of power laws as a model of opacity.

We adopted the same dust composition as Paper~I: two populations of large grains, one made of amorphous carbon accounting for 30\% of the dust mass, and one of amorphous magnesium-iron silicates making up 70\% of the mass. These abundances are consistent with typical Milky Way dust models \citep[e.g.,][]{Wein+Draine_01, Compiegne+11}. Our choice of carbon is the BE material from \citep{Mennella+98}, whose opacity was published in the form of a mass absorption coefficient $\kappa$. For silicates, we use the E30R material (Mg$_{0.7}$Fe$_{0.3}$SiO$_3$ with partially reduced iron) from \cite{Demyk+22}, whose optical properties were published as a complex refractive index ($n$, $k$) \citep[after having been originally published as $\kappa$ in][which is the data used in Paper~I]{Demyk+17B}.

\subsection{Treatment of the material opacity}
\label{sec:methodology:opacity}

We converted the optical properties of the two materials included in the model to mass absorption coefficients ($\kappa$) and determined a continuous interpolation between the experimentally measured temperatures (4 or 5 in the 20-300 K range).

For the optical properties of the E30R silicate, we used the interpolation method given in appendix C of \cite{Demyk+22}. Their table C.1 lists the parameters to calculate the temperature-dependent complex refractive index ($n,~k$) for all amorphous silicates considered by \cite{Demyk+17A,Demyk+17B}. The conversion of ($n,~k$) to a mass absorption coefficient depends on the shape and structure of the grains, so we had to choose a representative model for interstellar dust grains. Since we know that dust grains are non-spherical \citep[e.g.,][]{Das+10, Draine+Hensley_21}, we adopted compact, prolate (elongated) spheroids grains with an axial ratio of 2. One may question whether compact grains or aggregates are more appropriate in a galactic-scale dust model, since interstellar dust grains are expected to form fractal aggregates in dense galactic regions such as molecular clouds \citep[e.g.,][]{Ormel+09,Juvela+15}. Since this study focuses on temperature-driven variations in opacity rather than its absolute value, the fractal nature of grains is secondary and we defaulted to compact grains. Using the grain model thus defined, we calculated the silicate mass absorption coefficient using the formulae for spheroids in the Rayleigh regime from \cite{VdHulst57}. 

In the case of BE carbon, the published opacity is already in the form of a mass absorption coefficient $\kappa$. However, laboratory samples are typically made of grain aggregates rather than single particles, and aggregates have a higher value of $\kappa$ compared to compact grains of the same mass \citep[e.g.,][]{Ormel+11, Koehler+15, Ysard+18}. Since in our model we assumed that silicate grains are compact, we assume the same for carbonaceous grains, for consistency. Therefore, we applied a correction factor to the experimental carbon $\kappa$, to counteract the ``aggregation enhancement'' effect and estimate the mass absorption coefficient of compact BE grains. The modeling results of \cite{Ysard+18} for aromatic amorphous carbon show that this aggregation enhancement makes far-infrared opacity both higher and shallower, and can be approximated as a power law
(see their fig. 6):
\begin{align*}
    \frac{\kappa_{0,~\mathrm{aggregate}}}{\kappa_{0,~\mathrm{compact}}} = 2.6~\text{at 100 $\mu$m,} \\
    \beta_{\rm aggregate} - \beta_{\rm compact} = - 0.15
\end{align*}
In the rest of the paper, whenever we mention the carbon $\kappa$, we refer to the value thus corrected, unless stated otherwise (in Paper~I, which was focused on the coarser differences between experimental and theoretically extrapolated opacity, the aggregate enhancement was modeled as a constant factor). After correcting for the presence of aggregates, we interpolated carbon opacity over temperature. Overall, as shown in Appendix~\ref{sec:appendix_opacity_interpolation}, the opacity of BE carbon can be approximated as a power law where $\kappa_0$ and $\beta$ are quadratic functions of temperature.

\begin{figure}
\begin{center}
\includegraphics[width=.99\columnwidth]{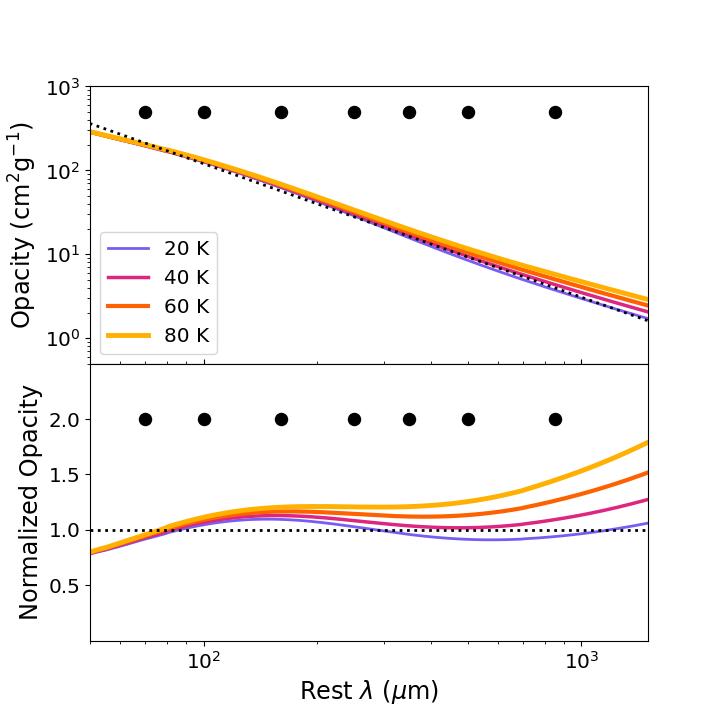}
\caption{Temperature dependence of dust opacity (colored curves) for our standard dust composition of 30\% BE carbon and 70\% E30R silicate, compared to a power law (dotted black line). The black symbols indicate the central wavelengths of the seven standard bands for $z$~=~0 (see Section~\ref{sec:methodology:fit}).
}
\label{fig:opac_vs_T}
\end{center}
\end{figure}

After obtaining $\kappa$ for both silicate and carbon in their final form, we calculated the overall value of $\kappa$ for our dust model as the sum of the opacity of the silicate grain population and the carbon population, weighted by the mass fraction of each component. For our chosen dust composition, this corresponds to  0.7 for silicates and 0.3 for carbonaceous materials (see the beginning of Section~\ref{sec:methodology}). The resulting $\kappa(\lambda)$, and its variation with temperature between 20 and 80~K, are plotted in Fig.~\ref{fig:opac_vs_T}. We also show a power law fit to the 20~K opacity to underline the difference between power law and realistic opacity: the experimentally-derived $\kappa$ has a ``knee'' at $\lambda \sim 150~\mu{\rm m}$ and an ``ankle'' at $\lambda \sim 600~\mu{\rm m}$, which are E30R silicate features.

The main changes in dust opacity with increasing temperature are the decrease in the average slope and the flattening of the knee and ankle features. The first of these two effects can be quantified by fitting the opacity with a power law, thus obtaining the value of $\kappa_0$ and an average value of the slope, $\beta_{\rm avg}$. The evolution of $\beta_{\rm avg}$ and $\kappa_0$ with temperature is shown in Table~\ref{tab:opac_vs_T} for $20\, {\rm K} \leq T \leq 250\, {\rm K}$ and for three different choices of $\lambda_0$. The variation of $\kappa_0$ with temperature is strongly dependent on the choice of the reference wavelength $\lambda_0$, with longer wavelengths showing greater variation. For $\lambda_0~=~500~\mu{\rm m}$, a typical value used in the literature, $\kappa_0$ varies by a factor $\sim2.5$ between 20 and 250~K. For $\lambda_0~=~850~\mu{\rm m}$, the value used in Paper~I, the $\kappa_0$ variation increases to a factor $\sim3$. On the other hand, for $\lambda_0~=~100~\mu{\rm m}$, $\kappa_0$ changes by less than a quarter in the same temperature range. Overall, the temperature dependence shown in Table~\ref{tab:opac_vs_T} can be modeled as a power law with a variable $\beta$ and a fixed or slowly changing $\kappa_0$, if one chooses $\lambda_0 \lesssim 100~\mu{\rm m}$. This description of the opacity evolution with temperature, of course, does not recapture one feature of the experimental $\kappa(\lambda)$, i.e. the presence of the knee and ankle features. The impact of neglecting these departures from a power law is explored in Section~\ref{sec:results:wl-sampling}.

\begin{table}
\centering
\caption{Parametrization of the opacity evolution with temperature: values of $\kappa_0$ (in cm$^2$g$^{-1}$) and $\beta_{\rm avg}$ for different values of temperature and $\lambda_0$. This power law model is a first-order approximation neglecting the effects of broad (``knee'' and ``ankle'') opacity features.}
\begin{tabular}{ccccc}
    \hline
    \textit{T}  & $\beta_{\rm avg}$ & $\kappa_{\rm 100~\mu m}$ & $\kappa_{\rm 500~\mu m}$ & $\kappa_{\rm 850~\mu m}$ \\
    (K) &  & (cm$^2$g$^{-1}$) & (cm$^2$g$^{-1}$) & (cm$^2$g$^{-1}$) \\
    \hline\hline
    20 & 1.59 & 120 & 9.27 & 3.99 \\
    50 & 1.50 & 123 & 10.9 & 4.93 \\
    100 & 1.37 & 128 & 14.1 & 6.80 \\
    150 & 1.27 & 134 & 17.3 & 8.81 \\
    200 & 1.21 & 142 & 20.2 & 10.6 \\
    250 & 1.17 & 148 & 22.3 & 11.9 \\
    \hline
\end{tabular}
\label{tab:opac_vs_T}
\end{table}

\subsection{Temperature distribution}
\label{sec:methodology:t_dist}

Aside from the temperature-dependent opacity $\kappa(\lambda,T)$, the other ingredient needed to calculate the dust emission SED is the dust temperature (see equation~\ref{eq_mbb}). However, a single-temperature model is not realistic when modeling a diverse environment such as a galaxy. Therefore, we constructed a more realistic model for dust temperature distribution. We adopted two simplifying assumptions, as we did in Paper~I, to reduce the complexity of the model needed. First, we assumed that carbon and silicate grains have the same temperature: while it is generally expected that carbonaceous grains will be warmer than silicates in the same environment, the expected difference is a few K \citep[e.g.,][Ch. 24]{Draine_book}, which is small compared to the width of the temperature distribution we adopt (see Section~\ref{sec:methodology:synthphot}). Secondly, we modeled our objects as optically thin: the influence of this simplification on the temperature distribution is explored later in this section.

Based on theoretical considerations on radiative transfer, \cite{Dale+01} argued that the interstellar radiation field intensity follows a power law with an exponent $\alpha$ which varies between 1 and 2.5, and is higher in more optically thin media. The seminal \citet{DL07} dust model adds a second component: a delta function at the minimum interstellar radiation field intensity, representing a population of dust illuminated by a uniform, moderate-intensity radiation field in the diffuse ISM. In this scenario, the power law component is taken to represent dust in highly irradiated environments, such as photodissociation regions (PDRs). The mass fraction \fpdr\ of PDR dust is typically a few percent in normal star-forming galaxies \citep[see, e.g., tables 4 and 5 of the SINGS galaxies study by][]{Draine+07}.
The temperature distribution of these PDR regions can be adequately modeled as a power law, as shown by \cite{Kovacs+10}. If the intensity of the radiation field is distributed as a power law of index $\alpha$ \cite[as per][]{Dale+01}, and dust opacity as a function of wavelength is a power law of index $\beta$ (equation~\ref{eq_kappa}), the dust temperature is also distributed as a power law of index $-s$, where $s = 2 + \alpha + \beta_{\rm eff}$ and $\beta_{\rm eff}$ is an effective opacity-corrected value of $\beta$ ($\beta_{\rm eff} = \beta$ in the optically thin limit). \cite{Kovacs+10} predicted a theoretical values of $s$ ranging from 4~--~5 in optically thick environments to 6.5~--~7.5 in the optically thin case.
Since dust properties change with temperature and do not necessarily have a constant slope $\beta$, this is a first-order approximation; however, a fully self-consistent model of dust temperature is beyond the scope of this paper, and we adopt this power law temperature distribution as a useful toy model.

We therefore modeled the temperature distribution as the sum of a delta function at $T~=~$\tmin, for diffuse ISM dust, and a power law at \tmin $<~T~<$ \tmax\ for PDR dust:
\begin{equation} \label{eq_Tdist}
\frac{dM_{\rm dust}}{dT} = (1 - f_{\rm PDR}) M_{\rm dust} \, \delta(T - T_{\rm min}) + f_{\rm PDR} C_{\rm norm} M_{\rm dust} \, T^{-s} 
\end{equation}
where \tmin\ and \tmax\ are the extremes of our temperature distribution, the power law exponent $s$ and the PDR dust fraction $f_{\rm PDR}$ have been defined above, and the normalization factor for the power law distribution is $C_{\rm norm} \equiv(s-1) / \left(T^{1-s}_{\rm min} - T^{1-s}_{\rm max}\right)$. This normalization requires that $s \neq 1$, which is satisfied in any physically plausible scenarios. The two extremes of this formula (\fpdr~=~0 or 1) correspond to a pure delta distribution (i.e., a single dust temperature for the whole galaxy) and a pure power law distribution, respectively. These two extremes can be interpreted as toy models for quiescent galaxies and a starburst, with intermediate values of the parameter representing intermediate star-forming rates.

To calculate the SED for this temperature distribution, we approximated it by a discrete series and expressed the total emission as a sum of $n_{\rm T}$ single-temperature modified blackbodies. The discretized version of equation~\ref{eq_Tdist} is:
\begin{equation} \label{eq_Tdist_discrete}
    \begin{aligned}
        T_{\rm i} &= T_{\rm min} + i \cdot \delta T\\
        \xi_{\rm i} &= \left\{
            \begin{aligned}
               & 1 - f_{\rm PDR} + \frac{(T_{\rm min})^{-s}}{\sum_{i=0}^{n_{\rm T}}(T_{\rm i})^{-s}} \cdot f_{\rm PDR} \quad & {\rm if~i = 0;} \\
                & \frac{(T_{\rm i})^{-s}}{\sum_{i=0}^{n_{\rm T}}(T_{\rm i})^{-s}} \cdot f_{\rm PDR} \quad & {\rm otherwise}.
            \end{aligned}
        \right.
    \end{aligned}
\end{equation}
where $T_{\rm i}$ is the i-th temperature value, $\delta T$ is the (fixed) temperature step,  $\xi_{\rm i}$ is the mass fraction of dust at temperature $T_{\rm i}$, and $\sum_{{\rm i} = 0}^{n_{\rm T}}\xi_{\rm i} = 1$. The sum is zero-indexed, with $T_0 =$~\tmin. The $i = 0$ bin is the only one to include a contribution from the delta distribution in addition to the power law. The choice of the temperature step is important, since numerical experiments show that low temperature resolution leads to an underestimate of short-wavelength emission. We chose a temperature step of 0.5~K as a compromise between precision and computational speed, since increasing the resolution further has rapidly diminishing returns: halving the step size from $\delta T = 1$~K to $\delta T = 0.5$~K increases the flux at $\lambda = 50\,\mu$m by $\sim 8 \%$, while halving it again to $\delta T = 0.25$~K only results in an increase of $\sim 4\%$.

\subsubsection{Mass-weighted temperature}
\label{sec:methodology:t_dist:t_mw}

To test the correctness of the fit, we are interested in comparing the derived temperature \tfit~to the physical temperature of dust. However, when multiple temperature components are present (as is the general case in the previous section), it is necessary to define the quantity used to represent the dust temperature distribution. Several such definitions have been devised over time \citep[see][for an overview]{Liang+19}. In this work, we adopted the mass-weighted temperature \tmw\ following \cite{Liang+19}, as the physical dust temperature to compare with \tfit. 

For the temperature distribution described by equation~\ref{eq_Tdist}, assuming no upper limit, the mass-weighted temperature can be calculated analytically and corresponds to:
\begin{equation}
\label{eq_tmw}
    {\rm T}_{\rm mw} = {\rm T}_{\rm min} \left( 1 + \frac{{\rm f}_{\rm PDR}}{s-2} \right)
\end{equation}
For \fpdr~=~0 (i.e., in the case of uniform dust temperature), this equation reduces trivially to \tmw~=~\tmin. For \fpdr~$\neq 0$, the value of \tmw\ depends on $s$: for $s = 5$ (optically thick medium) one has \tmw~$\sim$~\tmin~$\cdot$ (1 + 0.33~\fpdr), while for $s = 7.5$ (optically thin), \tmw~$\sim$~\tmin~$\cdot$ (1 + 0.18~\fpdr).

The precise expression for \tmw\ becomes more complex if one truncates the temperature distribution at a certain \tmax. However, equation~\ref{eq_tmw} remains accurate to better than $\simeq1\%$ if $s \geq 6$ and \tmax~>~3~\tmin, as shown in Appendix~\ref{sec:appendix_TMW}. Since these conditions are satisfied for all the models we used in our synthetic observation grid (see Section~\ref{sec:methodology:synthphot}), we decided to use equation~\ref{eq_tmw} without further corrections.

\subsection{Synthetic SED and photometry production}
\label{sec:methodology:synthphot}

To calculate the SED for a multi-temperature object, we made the simplifying assumption that it is optically thin at the wavelengths we are interested in ($\lambda > 50\,\mu$m). In this case, the total SED is simply the sum of the individual single-temperature components. The thermal emission can then be obtained by combining equations~\ref{eq_mbb} and \ref{eq_Tdist_discrete}:
\begin{equation}
\label{eq_mbb_multiple}
    F_\nu (\lambda) = \frac{1+z}{D_L^2}\, M_{\rm dust} \sum_{i=0}^{n_{\rm T}} \xi_i \ B_{\rm \nu}\left(\lambda_{\rm rest}, T_{i} \right) \kappa\left( \lambda_{\rm rest}, T_i \right) C_{\rm i,CMB}
\end{equation}
where $\kappa( \lambda_{\rm rest}, T_{\rm i})$ is the wavelength- and temperature-dependent dust opacity as described in Section~\ref{sec:methodology:opacity}, $\lambda_{\rm rest} = \lambda/(1+z)$ is the rest wavelength, and $C_{\rm i,CMB}$ is the CMB correction factor from \cite{daCunha+13} for temperature $T_{\rm i}$. 

\begin{figure}
\begin{minipage}{0.5\textwidth}
\begin{center}
\includegraphics[width=.99\hsize]{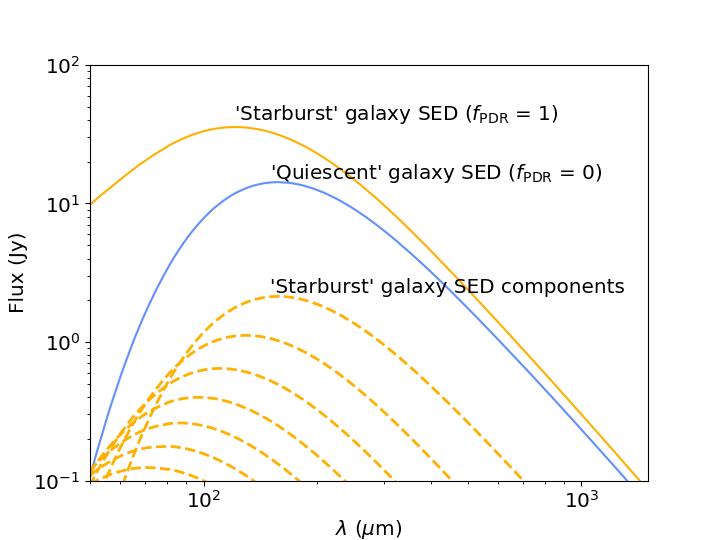}
\caption{
Different shapes of the SED for \fpdr~=~0 (single-temperature dust; blue line) and \fpdr~=~1 (power law temperature distribution; orange line). Dashed orange lines are individual temperature components for the SED with \fpdr~=~1. Only one every ten dashed lines is plotted, for legibility. Both SEDs use \tmin~=~20~K; the multi-temperature SED uses $s = 7.5$.
}
\label{fig:SED_model_example}
\end{center}
\end{minipage}
\end{figure}

\begin{table}
\centering
\caption{
Table of filter bands used to  produce multi-band photometry, with uncertainties, from a synthetic SED. The uncertainties on \textit{Herschel} and SCUBA-2 are the bands' confusion limit, while the uncertainties on ALMA are the rms noise for 2 hours of integration, which is why they are in different units. Sources for the uncertainties: for \textit{Herschel} (PACS and SPIRE) bands, \textit{Herschel} Observers' Manual, Section 4.3; 
for SCUBA-2 bands, \citet{Dempsey+13}; for ALMA bands, the sensitivity for 2 hours of integration from the ALMA sensitivity calculator (see main text).}
\begin{tabular}{lrr}
    \hline
Band & Central $\lambda$ & Uncertainty \\
     & $\mu$m            & $\mu$Jy/Beam   \\
    \hline\hline
    PACS70      & 70.0   & 100 \\
    PACS100     & 100.0  & 300 \\
    PACS160     & 160.0  & 1000 \\
    SPIRE250    & 250.0  & 6000 \\
    SPIRE350    & 350.0  & 6000 \\
    SPIRE500    & 500.0  & 7000 \\
    SCUBA2\_450 & 450.0  & 500 \\
    SCUBA2\_850 & 850.0  & 700 \\
    \hline
                &        &  $\mu$Jy \\
    \hline
    ALMA 10     & 342.6  & 470.0  \\
    ALMA 9      & 441.5  & 191.0  \\
    ALMA 8      & 740.2  & 37.9 \\
    ALMA 7      & 872.8  & 20.4 \\
    ALMA 6      & 1287   & 11.0 \\
    ALMA 5      & 1477   & 11.0 \\
    ALMA 4      & 2068   & 8.3 \\ 
    ALMA 3      & 3075   & 7.6 \\ 
    ALMA 1      & 7460   & 5.5 \\ 
    \hline
\end{tabular}
\label{tab:filters}
\end{table}

The result of equation~\ref{eq_mbb_multiple} is a synthetic spectrum of continuum dust emission. To convert it into multi-band photometry, which is the form most often used in SED fits, we convolved the spectrum with various filter profiles. We used a total of 17 filters from \textit{Herschel} (PACS and SPIRE, obtained from the SVO filter Profile Service \citep{SVO_ref_2012, SVO_ref_2020, SVO_ref_2024}), SCUBA-2 (from the \textit{Filters} page on the JCMT website\footnote{\url{https://www.eaobservatory.org/jcmt/instrumentation/continuum/scuba-2/filters/}}) and ALMA (from the ALMA Cycle 11 technical handbook). The full complement of filters covers the wavelength range between 70~$\mu$m and 7.5 mm (see Table~\ref{tab:filters}). Because ALMA allows flexibility in the choice of wavelength to observe, in the present work we calculated photometry at the nominal central wavelength for each band (Table~\ref{tab:filters}) and a total bandwidth of 7.5 GHz (four spectral windows of 1.875 GHz each). 
To construct a realistic observational uncertainty for \textit{Herschel} and SCUBA-2 synthetic photometry we used the sum in quadrature of calibration uncertainty and confusion limit. We adopted a calibration uncertainty of 7\% for PACS \citep{Balog+14}, 5.5\% for SPIRE \citep{Bendo+13}, and 16\% and 6.5\% for SCUBA-2 at 450 and 850~$\mu$m \citep{Mairs+21}; confusion limits are given in  Table~\ref{tab:filters}. For ALMA bands we assumed uncertainties to be either 10\% of the flux or the sensitivity for a 2-hour observation from the ALMA sensitivity calculator,\footnote{\url{https://almascience.eso.org/proposing/sensitivity-calculator}} whichever is greater. 

\begin{table}
\centering
\caption{
Parameter values used in the creation of our model grid (see main text).
}
\begin{tabular}{ll}
    \hline
    Parameter   & Values \\
    \hline\hline
    \tmin\ (K)  & 20, 25, 30, 35, 40, 50, 60, 70, 80 \\
    \fpdr\      & 0, 0.003, 0.01, 0.03, 0.1, 0.3, 1.0 \\
    $s$         & 6.5, 7.0, 7.5, 8.0 \\
    $z$         & 0, 0.25, 0.50, \ldots, 11.75, 12.00 \\
    \hline
\end{tabular}
\label{tab:param_grid}
\end{table}

We create a grid of models spanning a wide range of dust temperature distributions and redshifts. We use nine values of \tmin~ranging between 20 K (a typical temperature for a quiescent galaxy) and 80 K \citep[close to the temperature of the most extreme starbursts; see][]{Witstok+23}. We used a set of seven \fpdr~values covering the space between 0 and 1, where 0 represents a quiescent galaxy and higher values correspond to higher star formation rates. The expected value of the temperature power-law index $s$ for dust with $\beta \sim 1.5$ in the optically thin case is $\sim 7.5$, so we let this parameter span the $6.5 \leq s \leq 8$ range in steps of 0.5. Finally, we let the redshift $z$ vary between 0 and 12 in steps of 0.25. We calculated luminosity distances from $z$ assuming a $\Lambda$CDM cosmology with parameters ${\rm H}_0 = 70 {\rm km}\,{\rm s}^{-1}\,{\rm Mpc}^{-1}$, $\Omega_{\rm M} = 0.3$, and a CMB temperature of 2.725~K at $z$~=~0, as in Paper~I. For $z$~=~0 we set the distance to 100 Mpc. We rejected as unphysical any model where \tmin\ is lower than the CMB temperature. 
The full set of parameters is shown in Table~\ref{tab:param_grid}. Parameter values in common to all synthetic observations are the dust mass \minput\ of $10^8$ solar masses and the maximum temperature \tmax~$=$~250~K. The condition \tmax~>~3~\tmin\ for adopting equation~\ref{eq_tmw} as an approximation of \tmw\ is always satisfied.

The final result of the procedure is a grid of 9617 points, each representing an unresolved galaxy observation, which cover the entire span of physically realistic dust temperature distributions for $z \leq 12$. For each grid point we have calculated a far-infrared synthetic SED, plus synthetic far-infrared multi-band photometry. An example of SEDs in our model grid is shown in Fig.~\ref{fig:SED_model_example}, which compares the emission of two galaxies with the same value of \tmin\ (20~K), but different values of \fpdr\ (0 and 1, corresponding to a quiescent galaxy and an extreme starburst).

\subsection{Synthetic photometry fit}
\label{sec:methodology:fit}

We fit the synthetic photometry in our grid with single-temperature modified blackbody model (Eq.~\ref{eq_mbb}) with a power law opacity (Eq~\ref{eq_kappa}). Any differences between the fit results and the values used in the creation of the synthetic photometry provide a measure of the bias in the fit model and procedure. 

For the dust opacity to fit the synthetic observations, we chose parameters that most closely approximate the experimental $\kappa(\lambda)$ at $T= 20\,K$:  $\beta = 1.59$, $\kappa_0 = 120\, {\rm cm}^2 {\rm g}^{-2}$ and $\lambda_0 = 100\, \mu$m (Table~\ref{tab:opac_vs_T}). 
The choice of $\beta$ is close to the value of 1.5 used in Paper~I. The value of $\kappa_0$, on the other hand, corresponds to $\kappa(850\,\mu{\rm m}) \sim 4\, {\rm cm}^2 {\rm g}^{-1}$, significantly higher than the $\kappa(850\,\mu{\rm m}) = 0.7\, {\rm cm}^2 {\rm g}^{-1}$ used in Paper~I. We adopted the experimentally-derived value of $\kappa_0$, higher than the legacy value used in Paper~I, because this study focuses on the effect of opacity variation, while Paper~I focused on the average difference between experimental and legacy opacity.
Adopting these modified blackbody parameters is equivalent to imposing that the fitted mass \mfit\ is equal to the input value \minput\ for 20~K (single-temperature) dust, by construction. Therefore, in the discussion of fit results in Section~\ref{sec:results}, the goal is to understand what causes variations in \mfit\ rather than finding the combination of factors for which \mfit~=~\minput.

Each synthetic photometric set includes 17 bands; however, not all bands are necessarily useful: some may have a low signal-to-noise ratio (S/N), or be outside of the range of thermal dust emission if the synthetic source is at high redshift. Also, since real observations often have few observed bands -- especially at high $z$ -- we should limit the number of bands used for the fit, for the sake of realism. We aimed to retain at least the minimum number of bands for a well-determined fit, i.e., 3 for mass-temperature fits and 4 for mass-temperature-$\beta$ fits. On the other hand, we decided to limit each fit to a maximum of 7 bands for $z = 0$ sources, and 4 bands for sources with $z \geq 0.25$ (which is the smallest $z > 0$ value in our grid). To fulfill these conditions, we applied the following selection to our synthetic photometry. First of all, we excluded any band with a S/N below 3 (as calculated from the synthetic flux and the uncertainties calculated in Section~\ref{sec:methodology:synthphot}). Secondly, we threw out any band with a rest wavelength $\lambda_{\rm rest} = \lambda$/(1+z) shorter than 50~$\mu$m since, below this wavelength, galactic SEDs contain significant emission from stochastically-heated grains \citep[e.g.,][]{Casey_12}. After this selection, we excluded the synthetic galaxies left with no bands at $\lambda_{\rm rest} \leq 160\, \mu {\rm m}$, to avoid badly constrained dust temperatures. Thirdly, we excluded band pairs that cover similar wavelengths, since they don't provide as much information. We did this by excluding ALMA band 10 (9, 7) if the photometry included the \textit{Herschel} SPIRE 350~$\mu$m band (SCUBA-2 450-$\mu$m band, SCUBA-2 850-$\mu$m band). 
Finally, if after the previous step the number of remaining bands was larger than 7 (for $z = 0$) or 4 (for $z \geq 0.25$), we only retained the first 7 (4) bands in the following order of choosing: PACS bands, SPIRE bands, SCUBA-2 850~$\mu$m, SCUBA-2 450~$\mu$m, and ALMA 7, 6, 3, 9, 1, 8, 5, 4 and 10. This is to simulate the typical use frequency of different bands: many galaxies have been observed by \textit{Herschel}, and ALMA bands 3, 6 and 7 are more used than band 10.\footnote{For instance, bands 3, 6 and 7 are often used in ALMA Large Programs at high redshift, such as ASPECS \citep[][bands 3 and 6]{ALMA_ASPECS}; ALPINE \citep[][band 7]{ALMA_ALPINE} and REBELS \citep[][bands 6 and 7]{ALMA_REBELS}.}

We fit the photometric bands thus selected with the $\chi^2$-minimizing function \texttt{curve\_fit} from the Python package \texttt{SciPy} \citep{scipy}. We fitted for \mfit, \tfit, and (where available) \bfit, and also obtained the corresponding standard deviations from the covariance matrix to use them as the parameters' uncertainties. The fit includes the effect of filter convolution and CMB correction.  For a small fraction of the synthetic photometry samples (62 out of 9617) the fitting method did not converge to a solution. This happened in high-redshift synthetic sources where the CMB temperature is close to \tmin. We excluded these sources from our analysis. 

\section{Results}
\label{sec:results}

For the sake of conciseness, we discuss the results for a single value of the temperature distribution slope: $s = 7.5$. The effect of alternative $s$ values is explored in Appendix \ref{sec:appendix_s_effect}.

\subsection{Fixed $\beta$ fits}
\label{sec:results:fixbeta}

\begin{figure}
\begin{minipage}{0.5\textwidth}
\begin{center}
\includegraphics[width=.99\hsize]{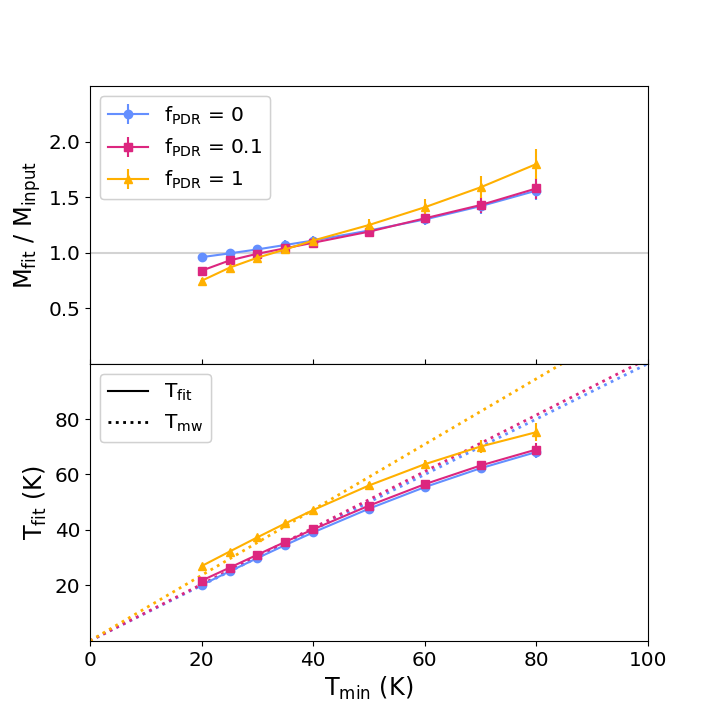}
\caption{
Results of the fixed-$\beta$ fits. \textbf{Top panel:} \mfit~normalized by \minput. The horizontal grey line corresponds to \mfit~=~\minput. \textbf{Bottom panel:} \tfit~(bottom) as a function of \tmin. The dotted lines show \tmw\ as a function of \tmin\ for \fpdr~=~0, 0.1 and 1. In the case \fpdr~=~0 (blue dotted line), \tmw~=~\tmin. Error bars show the parameter's standard deviation, from the fit.
}
\label{fig:fit_res_fixbeta}
\end{center}
\end{minipage}
\end{figure}

For a fit with a fixed $\beta$ value, the derived mass \mfit\ shows a bias that depends on both the minimum temperature \tmin\ and the shape of the dust temperature distribution as parametrized by the PDR mass fraction \fpdr. This is evident in Fig.~\ref{fig:fit_res_fixbeta}, which illustrates the case  $z = 0$ for a subset of the grid (each point corresponds to a single grid point, or SED fit result). The synthetic photometry used for the fit includes seven bands: all PACS and SPIRE bands, plus SCUBA-2 at 850 $\mu$m. We show the results for the full \tmin\ range 20--80~K, and for three values of \fpdr: 0 (corresponding to a quiescent, uniform-temperature galaxy), 0.1 (moderately high star formation rate), and 1 (starburst with power law $T$ distribution).
All synthetic observations use the same value of the input dust mass \minput\ ($10^8\, M_\odot$), so that discussing \mfit\ is equivalent to discussing the \mfit/\minput\ ratio. 

We observe several biases in the fit results. First of all, the value of \mfit\ increases with \tmin. This bias depends on the temperature distribution and is stronger for larger values of \fpdr. For quiescent galaxies (\fpdr~=~0), \mfit\ is almost unbiased at \tmin~=~20~K and increases by a factor 1.6 at 80~K. For an extreme starburst (\fpdr~=~1), the fitted mass is underestimated at 20~K (\mfit\ $\sim 0.75$ \minput) and increases by a factor 2.4 at 80~K. Since this increase is approximately linear with \tmin, we can use a linear fit to determine the slope of these trends. In quiescent galaxies, \mfit(\tmin) / \mfit(20 K) increases by $\sim 1.1 \cdot 10^{-2}$ per K above \tmin~=~20~K; in starburst galaxies, the slope is $\sim 2.2 \cdot 10^{-2}$ per K.

The comparison between the derived temperature \tfit\ and the mass-averaged temperature \tmw\ (Fig.~\ref{fig:fit_res_fixbeta}, bottom panel) shows the opposite bias: temperatures are well-recovered (\fpdr $\leq 0.1$) or slightly overestimated (\fpdr~=~1) at low \tmin, but they are underestimated at high \tmin. 

These biases on \mfit\ and \tfit\ have two main causes: the variation of dust opacity with temperature, and the presence of multiple temperature components on the line of sight. The slope of the dust far-infrared opacity becomes less steep as dust becomes warmer (see Fig.~\ref{fig:opac_vs_T}): if we approximate opacity as a power law, this means that $\beta$ decreases as $T$ increases. Since the current fit uses a fixed value of $\beta$ calibrated on $T = 20\, {\rm K}$ dust, it overestimates $\beta$ at high temperature, which leads to an underestimated \tfit\ and an overestimated \mfit. This effect is responsible for the high value of \mfit\ at high \tmin, and is more evident for larger PDR fractions since these have more hot ($T \gg$~\tmin) dust. 
On the other hand, multiple temperature components can also cause modified blackbody fits to \textit{under}estimate dust mass \citep{Shetty+09b}. This effect is of course seen when dust has a wide temperature distribution (\fpdr~>~0), and is more pronounced for higher values of \fpdr. It is also most pronounced when \tmin\ is low, i.e., when the temperature contrast between the cold and the warm ends of the temperature distribution is highest. Overall, both phenomena -- the temperature dependence of dust opacity and the presence of multiple temperature components -- result in an increase of the recovered mass \mfit\ for increasing \tmin. At low \tmin, the bias is mainly driven by the presence of multiple temperature components, while at high \tmin, it is mainly driven by the change in dust properties.

\subsection{Free $\beta$ fits}
\label{sec:results:free-beta}

\begin{figure}
\begin{minipage}{0.5\textwidth}
\begin{center}
\includegraphics[width=.99\hsize]{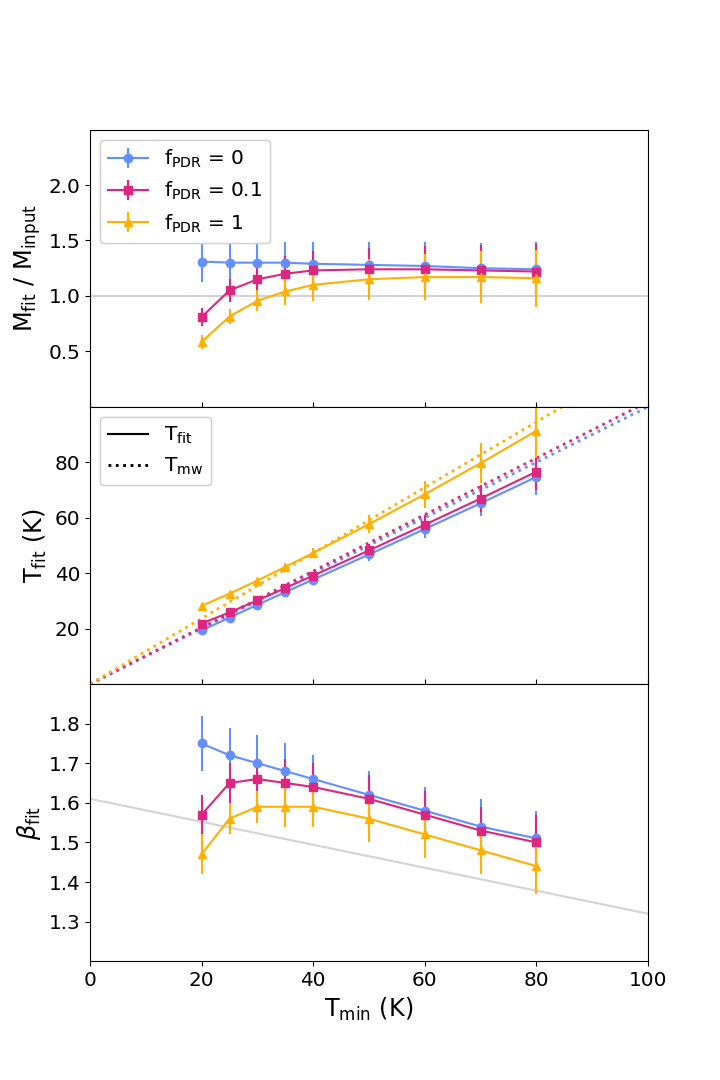}
\caption{
Results of the free-$\beta$ fits. \textbf{Top panel:} \mfit/\minput~as a function of \tmin. The horizontal grey line corresponds to \mfit~=~\minput. \textbf{Middle panel:} \tfit~as a function of \tmin. The dotted lines show \tmw~for the different values of \fpdr. \textbf{Bottom panel:} \bfit~as a function of \tmin. The grey line indicates the expected value of $\beta$ for $T =$~\tmin, interpolated from the values in Table~\ref{tab:opac_vs_T}.}
\label{fig:fit_res_freebeta}
\end{center}
\end{minipage}
\end{figure}

One of the main drivers of bias in the previous section is the change in the slope of opacity, which can be seen as the average $\beta$ of the model. It is natural then to ask if the fit gives more reliable results when $\beta$ is left as a free parameter. We repeated the fit on the same SEDs and bands as the previous section with a free $\beta$. The fit results, \mfit, \tfit\ and \bfit, are shown in Fig.~\ref{fig:fit_res_freebeta} for different star formation activities:\fpdr~=~0, 0.1, and 1.

The first obvious result is that for quiescent, uniform-temperature galaxies (i.e., \fpdr~=~0), the temperature dependence of \mfit\ is now flat, as shown in the top panel of Fig. \ref{fig:fit_res_freebeta}. Additionally, the mass value is consistently overestimated by $\sim 30\%$. This is likely due to the fact that the opacity is not a true power law, as we will explore more in depth in Section~\ref{sec:results:wl-sampling}. For wider temperature distributions (\fpdr~>~0), the presence of multiple temperatures on the line of sight affects fit results at low \tmin\ (see Section~\ref{sec:results:fixbeta}), so that \mfit\ increases up to \tmin~$\sim$~40~K, then flattens as it gets close to the value found for \fpdr~=~0. This is consistent with the idea that the temperature-dependent bias on \mfit, described in the previous section, is mainly due to a decrease of the opacity's slope (the ``average $\beta$'') with temperature. 

The middle panel in Fig.~\ref{fig:fit_res_freebeta} shows the recovered temperature \tfit\ as a function of \tmin. At high \tmin, the free-$\beta$ fit does a much better job of recovering a value close to the mass-weighted temperature \tmw\ (indicated by the dotted lines in Fig.~\ref{fig:fit_res_freebeta}), at the cost of larger error bars. For low \tmin, fit results are similar to the case of the fixed-$\beta$ fit: \tfit\ is close to \tmw\ in case of quiescent galaxies (\fpdr~=~0) while over-estimating it for actively star-forming galaxies (\fpdr~>~0). As in the previous section, the underestimate of \mfit\ and the overestimate of \tfit\ at low temperature are due to the presence of multiple temperature components on the line of sight.

Finally, the variation of \bfit\ is shown in the bottom panel of Fig.~\ref{fig:fit_res_freebeta}, where the fit results are compared to the intrinsic $T$-$\beta$ trend shown in Table~\ref{tab:opac_vs_T} (the gray line shows the expected $\beta$ for $T$~=~\tmin). 
The fit consistently overestimates $\beta$, likely as a consequence of the non-power-law opacity, with the exception of the models with \fpdr~>~0 and low \tmin where the the presence of multiple temperature components drives \bfit\ down. Despite this systematic shift, the fit results accurately recover the intrinsic anticorrelation between temperature and $\beta$: first, \bfit\ tends to decreases with \tmin, and secondly, even at fixed \tmin, \bfit\ is lower for higher \fpdr. This remains true even at \tmin $\gtrsim$ 40 K, where the presence of multiple temperature components on the line of sight has a negligible effect. 

In conclusion, adopting a variable $\beta$ fit where possible will greatly reduce the temperature-dependent bias caused by the variation of opacity with temperature. However, a temperature-independent bias remains. In the next section we explore this more in depth and find its likely origin in the fact that dust opacity is not a real power law. Finally, the effect of multiple temperature components on the line of sight remains, as can be expected in a single-temperature fit.

\subsection{Effect of wavelength sampling on non-power law opacity}
\label{sec:results:wl-sampling}

\begin{figure}
\begin{center}
\includegraphics[width=.99\columnwidth]{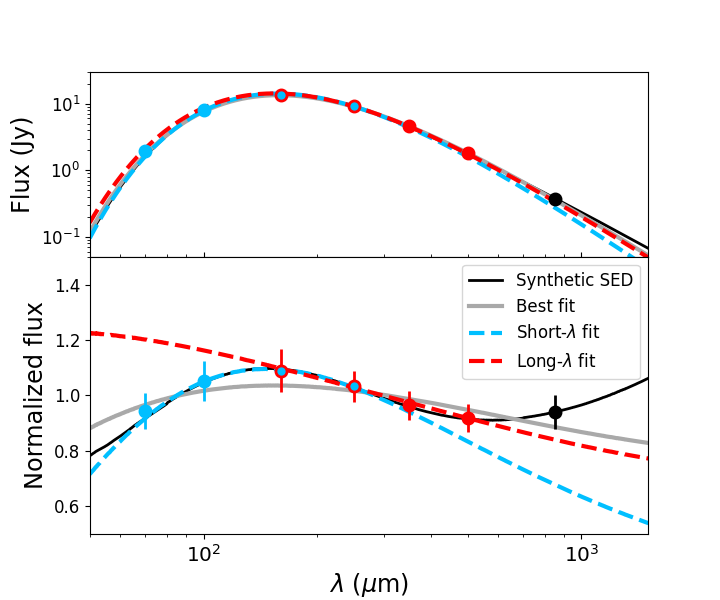}
\caption{Effect of the choice of bands on fit results. \textbf{Top panel:} A 20~K SED with non-power law opacity (black curve), photometry for 7 bands (colored dots), and three modified blackbody fits to the photometry (colored curves). Dots are color-coded depending on the fits they are used in. Dots used in two separate fits are bicolor. Color code: blue uses 4 bands between 70 and 250~$\mu$m; red uses 4 bands between 160 and 500~$\mu$m; grey uses all 7 bands (70 to 850~$\mu$m). \textbf{Bottom panel:} the same curves and photometry, normalized by a 20~K modified blackbody with $\beta = 1.59$ to highlight the different curve shapes, as well as the photometric error bars (which are not visible on the scale of the upper panel).}
\label{fig:band_dependent_fit_results_SEDs}
\end{center}
\end{figure}

As discussed in Section~\ref{sec:methodology:opacity}, the experimentally-derived opacity of our dust is not a perfect power law, but presents broad features which we called ``knee'' and ``ankle''. In this section, we show that neglecting the existence of these features introduces an almost temperature-independent bias, akin to the one seen in Section~\ref{sec:results:free-beta}. Notably, this bias depends on the (rest-frame) wavelength range of the fit, which in turn is determined by on the choice of bands and on the redshift of the source. As can be seen in Fig.~\ref{fig:band_dependent_fit_results_SEDs}, different wavelength ranges result in different effective opacities. For our choice of dust composition, opacity in the 70-250~$\mu$m range (blue and partly blue dots) shows a significant ``knee'', while in the 160-500~$\mu$m range (red and partly red dots) opacity is well approximated by a power law. This is in addition to the fact that opacity in both ranges changes with temperature. 

Fig.~\ref{fig:band_dependent_fit_results_SEDs} also shows the effect of the chosen wavelength range on a free-$\beta$ modified blackbody fit. We compare a synthetic SED from our grid (\tmin~=~20~K, \fpdr~=~0) with fits that use three different wavelength ranges: a ``best fit'' using all bands between 70 and 850~$\mu$m (i.e. the same bands as the fits in Section~\ref{sec:results:free-beta}; orange curve), one using the bands between 70 and 250~$\mu$m (blue curve) and one using the bands between 160 and 500~$\mu$m (red curve). Compared to the ``best fit'' curve, the 70-250~$\mu$m fit underpredicts emission at both long and short wavelength due to the aforementioned ``knee'' in the opacity, and it can be expected to return a lower \tfit\ and a higher \bfit. On the other hand, the 160-500~$\mu$m fit overpredicts short-wavelength emission but follows closely the ``best fit'' curve at long wavelengths, which would bias the fit results to a higher temperature and, therefore, a lower mass.

\begin{figure}
\begin{center}
\includegraphics[width=.99\columnwidth]{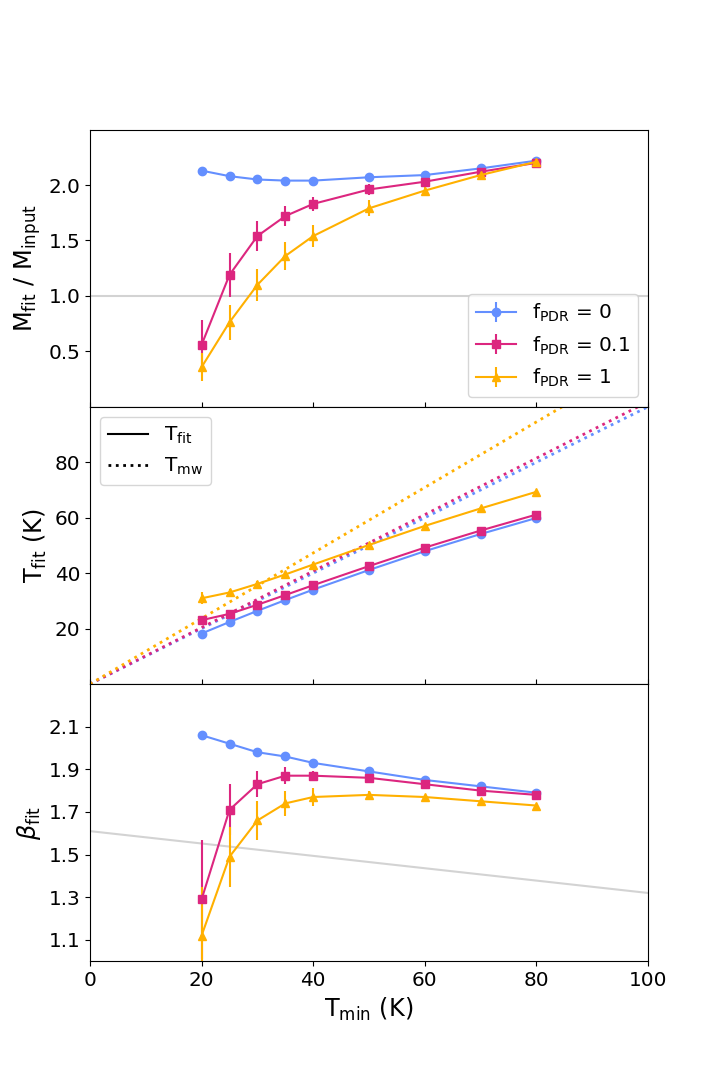}
\caption{
Results of a free-$\beta$ fit (see Fig.~\ref{fig:fit_res_freebeta}) when limited to the short-wavelength 70-250~$\mu$m range (blue data from Fig.~\ref{fig:band_dependent_fit_results_SEDs}). The color and line conventions are the same as in Fig.~\ref{fig:fit_res_freebeta}. The y-axis scale for the \bfit\ (bottom) plot has been changed to cover the full range of variation. 
}
\label{fig:fit_res_freebeta_shortwl}
\end{center}
\end{figure}

\begin{figure}
\begin{center}
\includegraphics[width=.99\columnwidth]{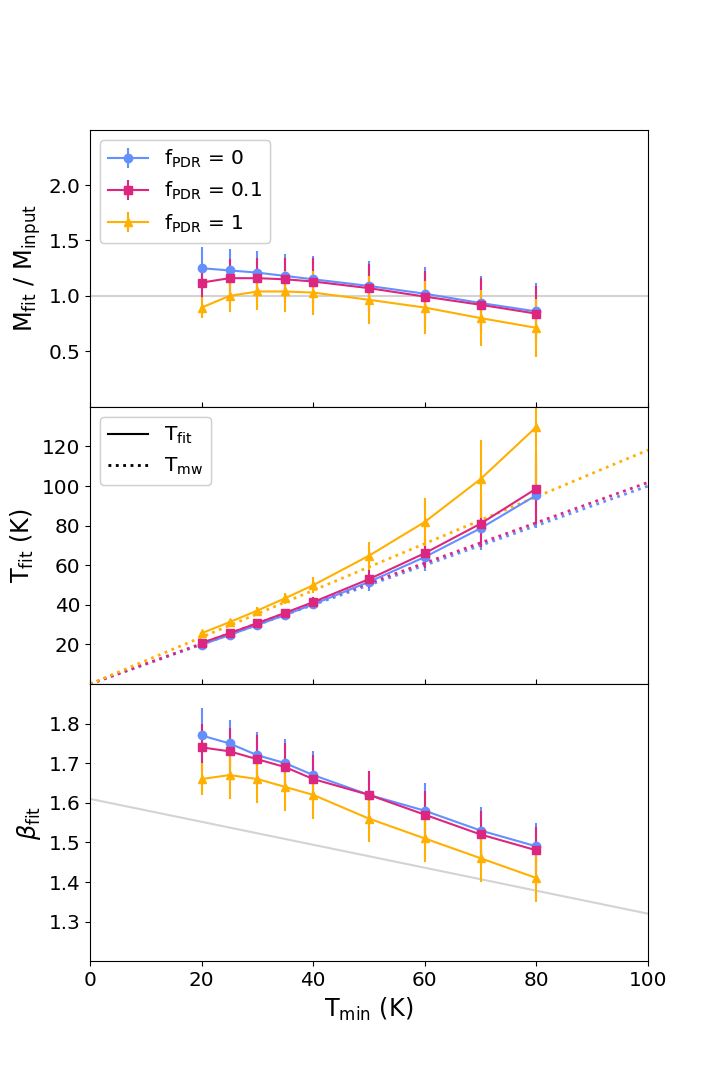}
\caption{
Results of a free-$\beta$ fit (see Fig.~\ref{fig:fit_res_freebeta}) when limited to the long-wavelength 160-500~$\mu$m range (red data from Fig.~\ref{fig:band_dependent_fit_results_SEDs}). The color and line conventions are the same as in Fig.~\ref{fig:fit_res_freebeta}. The y-axis scale for the \tfit\ (middle) plot has been changed to cover the full range of variation.
}
\label{fig:fit_res_freebeta_longwl}
\end{center}
\end{figure}

Because of these differences, the two fits give very different results for \mfit. The results for a wavelength range restricted to 70-250~$\mu$m (160-500~$\mu$m) are shown in Fig.~\ref{fig:fit_res_freebeta_shortwl} (Fig.~\ref{fig:fit_res_freebeta_longwl}).
For the 70-250~$\mu$m range, and for quiescent galaxies (\fpdr~=~0), the fit consistently returns \mfit$~\sim~2$~\minput\ (top panel). This high \mfit\ is consistent with the low \tfit\ caused by the ``knee'' (middle panel), as discussed previously. On the other hand, the SEDs corresponding to star-forming galaxies (\fpdr$~>~0$) are affected by the presence of multiple temperature components and have smaller derived dust masses at low \tmin, similar to what is observed in Fig.~\ref{fig:fit_res_freebeta}. In fact, the effect of multi-temperature dust is stronger here, since we are focusing on shorter wavelengths. The fit to the 160-500~$\mu$m range (Fig.~\ref{fig:fit_res_freebeta_longwl}) shows a closer match between \mfit\ and \minput, since opacity is closer to a power law in this wavelength range. The effect of multiple temperature components is also much smaller, due to the longer wavelengths used.

\begin{figure}
\begin{center}
\includegraphics[width=.99\columnwidth]{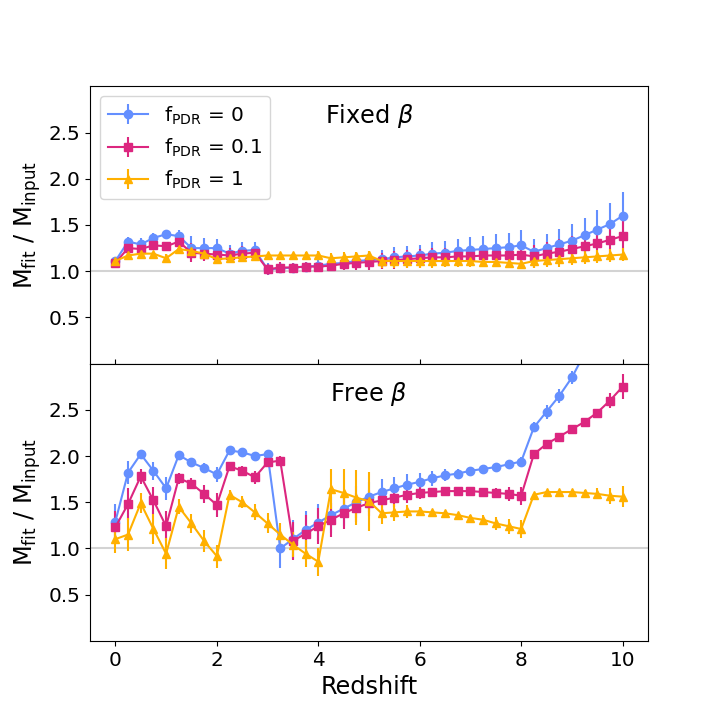}
\caption{Effect of redshift on the recovered value of dust mass, shown for \tmin~=~40~K, and for both fixed and free $\beta$ fits. Different redshifts correspond to different wavelength samplings and therefore different values of the bias on \mfit. The result is a significant scatter in \mfit/\minput. This is more notable for free-$\beta$ fits, as mentioned in Section~\ref{sec:results:free-beta}.}
\label{fig:Mfit_results_vs_z}
\end{center}
\end{figure}

This has important consequences on the comparison of mass fits for objects at different redshifts. The SED of sources at different $z$ will generally be sampled at different rest wavelengths. 
This is because different instruments and bands are effective in different redshift ranges, and even for objects observed in the same bands (e.g., as part of a same survey), the same observed wavelength corresponds to different rest wavelengths for objects at different redshifts. As shown above, when opacity is not a power law, different wavelength samplings will result in different fit results (most notably different \mfit) even when everything else is equal. To quantify this effect, we calculate \mfit/\minput\ for a galaxy with \tmin~=~40~K and for values of $z$ between 0 and 10, using the band selection described in Section~\ref{sec:methodology:fit}. For z~>~0, we move from the 7-band selection used in Section~\ref{sec:results:fixbeta} and \ref{sec:results:free-beta} to a 4-band selection, since high-redshift observations tend to have few(er) viable bands. While 4-band photometry of galaxies at high redshift is not common, it is also not unprecedented: for instance, \cite{Witstok+23} compile a list of 17 galaxies at 4 < $z$ < 8 observed in 4 or more photometric bands. The results of our 4-band fits are plotted in Fig.~\ref{fig:Mfit_results_vs_z}: overall, as $z$ changes, the variation in rest wavelength sampling introduces a scatter on the value of \mfit/\minput. The effect is modest, although measurable, for fixed-$\beta$ fits (top panel). However, it is much more important for free-$\beta$ fits, which shows \mfit/\minput\ variations by a factor of 2 or higher. This is consistent with our finding that free-$\beta$ fits are very sensitive to the effect of non-power law opacity (see Figs.~\ref{fig:band_dependent_fit_results_SEDs}, \ref{fig:fit_res_freebeta_shortwl} and \ref{fig:fit_res_freebeta_longwl}). Another result is that \mfit\ is generally overestimated, although the value of this bias is highly model-dependent. This bias is on average larger for free-$\beta$ fits, suggesting that it is due to a similar mechanism as the bias in Fig.~\ref{fig:fit_res_freebeta_shortwl}: most of our fits cover the emission peak, which is close to the opacity's ``knee''. A small contribution to this bias may also come from the fact that we are fitting galaxies at \tmin~=~40~K with our standard opacity calibrated on 20~K dust opacity. 

The change in fit results depends not only on the wavelength range selected, but on the spectral shape of opacity as well. For instance, the high \mfit\ value for the 70-250~$\mu$m range is due to the ``knee'' centered at about 150~$\mu$m, which is characteristic of the \cite{Demyk+22} silicate we used in our dust model. Different dust compositions may result in different positions of ``knees'' and/or ``ankles''.

\section{Astrophysical implications}
\label{sec:implications}

We have shown that dust masses derived from modified blackbody fits present a bias due to the temperature-dependent, non-power law nature of dust opacity. This bias depends on both dust temperature and (rest-frame) wavelength sampling. This has important consequences on the estimation of dust mass and temperature from modified blackbody fits, and their interpretation.

\begin{figure}
\begin{center}
\includegraphics[width=.99\columnwidth]{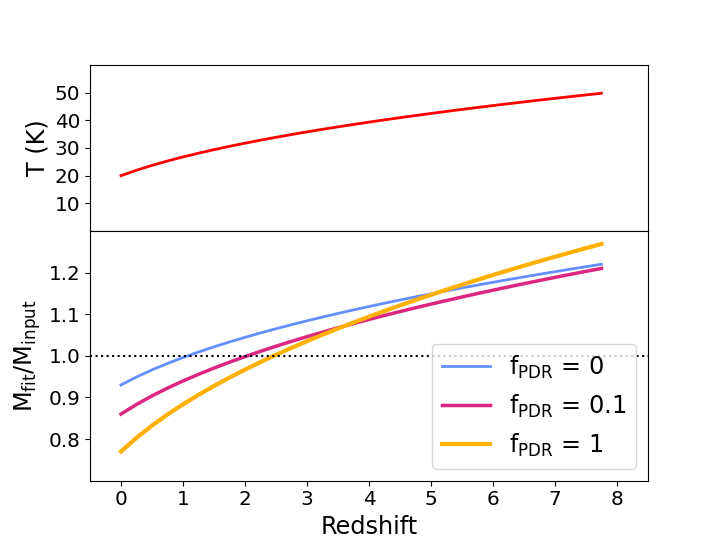}
\caption{Evolution of the expected value of \tmin\ (top) and \mfit/\minput\ (bottom) as a function of redshift, as predicted by Sommovigo's $T(z)$ formula (see text) and the linear \tmin -- \mfit/\minput\ relation found in Section~\ref{sec:results:fixbeta}.}
\label{fig:Mfit+T_vs_z_prediction}
\end{center}
\end{figure}

One field in which our findings have nontrivial consequences is the comparison of dust masses in the local Universe to masses at high redshift. While the specifics of dust temperature evolution with redshift are debated \citep[e.g., ][]{Liang+19, Sommovigo+22, Eales&Ward_24}, it is generally accepted that in the early universe the dust temperature was higher than today, due to a combination of higher star-formation rates and CMB heating. this is equivalent to \fpdr\ and \tmin increasing with $z$. Following from the results in the previous sections, this would imply that high-redshift dust masses are overestimated compared to local dust masses. To quantify this effect, we use the redshift-temperature dependence from \cite{Sommovigo+22}, $T \propto (1+z)^{0.42}$. 
If we use this $T(z)$ formula as a proxy for the \tmin\ of galaxy temperature distribution, and we assume \tmin~=~20~K as typical of the local Universe, \tmin\ evolves with redshift as shown in Fig.~\ref{fig:Mfit+T_vs_z_prediction} (top). Assuming that dust masses are obtained via a fixed $\beta$ fit, we can use the linear relation between \tmin\ and \mfit/\minput\ from Section~\ref{sec:results:fixbeta} to calculate the evolution of \mfit\ as a function of $z$. The result in shown in Fig.~\ref{fig:Mfit+T_vs_z_prediction} (bottom). For our model grid, the fit result \mfit/\minput\ increases by $\sim$25\% (\fpdr=0) to $\sim$60\% (\fpdr=1) between $z=0$ and 8. Therefore, among the sources of uncertainty in dust mass determinations at high redshift, the temperature dependence of dust opacity alone may cause a 25-60\% overestimate of dust masses at $z \sim 8$ compared to dust masses in the local universe. Unlike other sources of bias, this one can be estimated from $z$ or \tfit\ and corrected for.

The effects of temperature-dependent opacity do not apply solely to the high-redshift universe; they affect all systems with large temperature gradients. For instance, this uncertainty would apply to the comparison between a quiescent and a starburst galaxy, which has important implications for the study of the ISM in the most intense star-forming regions in the local Universe. Another issue affected by our finding is the estimate of dust formation around evolved stars. In the study of dust-bearing winds around AGB stars \citep[see e.g.][]{Thavisha+18}, warmer dust near the star would have higher opacity than the dust at larger distance. If not accounted for, this may distort the derived radial surface density profile.

Experimental measurements of temperature-dependent opacity can also help understand another long-standing astrophysical puzzle: the origin of the well-known anticorrelation between dust temperature and $\beta$. This anticorrelation has been well documented in astronomical observations \citep[e.g.,][and refs. therein]{Dale+Helou_02}, but it is not clear to what extent it is a physical phenomenon \citep{Boudet+05, Meny+07, Paradis+11} and to what extent it is observational bias caused by noisy data \citep{Shetty+09a, Shetty+09b}. Observational bias can be reduced, though not entirely eliminated, by using Bayesian methods or assuming a functional $\beta(T)$ form for the fit \citep{Juvela+13}. A physically-motivated, experimentally-measured $\beta(T)$ relation such as that presented in this work can provide helpful constraints on the Bayesian/functional methods used to recover the intrinsic $\beta$-$T$ anticorrelation.

Finally, we underline that these results depend on the shape of $\kappa(\lambda)$, and therefore on dust composition. The quantitative results in this study are tied to our choice of 30\% BE carbon + 70\% E30R silicate dust. However, the general trends in dust opacity ($\beta$ decreases with temperature, presence of a far-infrared ``knee'' in silicates) are found more robustly across species. Overall, different dust compositions are likely to result in different fit parameters, but in the same general trends between the derived mass \mfit\ and the dust temperature (either the minimum temperature \tmin\ or the mass-averaged temperature \tmw). It may be an interesting exercise to repeat the analysis for different dust compositions, to estimate a systematic uncertainty on \mfit/\minput\ from the dispersion of the results.

\section{Conclusions}
\label{sec:conclusions}

One of the most common methods for obtaining the mass and other properties of interstellar dust is fitting the dust emission SED using a modified blackbody model. However, far-infrared dust opacity is both temperature-dependent and not a true power law. 
In this work, we use the experimentally measured temperature-dependent opacity of dust analogues to construct the synthetic SED for a grid of dusty galaxies, which we then fit with standard modified blackbody models. 
We performed a comprehensive, experimentally-justified quantification of the bias caused by temperature-dependent, non-power law opacity.

Our main findings are the following:
\begin{itemize}
    \item The change of opacity with increasing temperature can be modeled as a decrease in its slope ($\beta$ in a power law approximation), while the reference opacity $\kappa_0 = \kappa(\lambda_0)$ is approximately constant, as long as the reference wavelength is short ($\lambda_0 \lesssim 100\,\mu {\rm m}$).
    \item The derived dust mass \mfit\ from fixed-$\beta$ fits is affected by the temperature dependence of far-infrared opacity. In the most favorable case (quiescent single-temperature galaxy), the derived dust mass increases almost linearly by $\sim$60\% between 20~K and 80~K. This is from temperature dependence alone, without considering other sources of systematics (e.g., uncertainty in the value of $\kappa_0$). 
    \item In galaxies with wider temperature distributions (e.g., starbursts with significant PDR contribution), the change in \mfit\ with temperature is larger: for the extreme cases in our model grid, the relative increase between 20~K and 80~K is over twice as large as in the single-temperature case. This is in part because dust in the high-temperature tail has higher opacity, and in part because there are multiple temperature component on the line of sight \citep{Shetty+09b}, which causes a bias in the same direction. 
    \item Where data allows it, using free-$\beta$ fits, rather than fixed-$\beta$, reduces or avoids this temperature-dependent bias. However, free-$\beta$ fits are more sensitive to the presence of ``knees'' or ``ankles'' in a non-power law $\kappa(\lambda)$. This introduces a different and potentially large bias, which is independent of temperature but dependent on the (rest-frame) wavelength range sampled.
    \item Both fixed-$\beta$ and free-$\beta$ fits are likely to introduce a bias when comparing dust mass measurements in the local Universe and at high redshift, since this involves the comparison of different temperature environments and different rest-frame wavelength sampling. 
    \item Our findings depend on the $\kappa(\lambda)$ profile used (i.e., the position and intensity of the ``knees'' and ``ankles''), meaning that the numerical results are characteristic of the dust composition we assumed in the present work. Alternative compositions should be studied to find the limits and uncertainties of our results. Nonetheless, since the opacities of most materials show qualitatively similar evolution with temperature, we do not expect dust composition to alter the main thrust of our findings.
\end{itemize}

\section*{Acknowledgements}
We are thankful to N.~Ysard for providing us with the data from fig.~6 of \cite{Ysard+18}. 
This paper significantly benefited from discussions with T.~Bakx and L.~Sommovigo. 
LF acknowledges support from the National Science and Technology Council of Taiwan under grants No. 111-2112-M-005-018-
MY3 and 114-2811-M-032-005.
JPM acknowledges support by the National Science and Technology Council of Taiwan under grant NSTC 112-2112-M-001-032-MY3.
FK acknowledges support from the Spanish Ministry of Science, Innovation and Universities, under project PID2023-149918NB-I00, financed by MCIU /AEI /10.13039/5
01100011033 / FEDER, EU.
SS acknowledges support from the UNAM-PAPIIT Program IA104824.
This work was also partly supported by the Spanish program Unidad de Excelencia María de Maeztu CEX2020-001058-M, financed by MCIN/AEI/10.13039/501100011033.
This research has made use of the SVO Filter Profile Service "Carlos Rodrigo", funded by MCIN/AEI/10.13039/501100011033/ through grant PID2023-146210NB-I00.

\section*{Data Availability}

The code and data for this paper can be found in the GitHub repository at \url{https://github.com/ICSM/Fanciullo_et_al_2026_temperature-dependent-opacity}, and as the Figshare item at the following DOI: \url{https://doi.org/10.6084/m9.figshare.32974043}.

This work made use of the following Python packages: \texttt{Numpy} \citep{numpy}, \texttt{SciPy} \citep{scipy}, \texttt{Astropy}\footnote{http://www.astropy.org} \citep{astropy_2013, astropy_2018, astropy_2022}, \texttt{Matplotlib} \citep{Matplotlib}, and \texttt{pandas} \citep{pandas_paper, pandas2.0.3}.

This paper uses of the IBM Design Language color blind friendly palette, with hex codes recovered from Mark Tucker's website.\footnote{https://apl.ninja/MarkTucker/color-blind-palettes-3inj}



\bibliographystyle{mnras}
\bibliography{Biblio} 



\appendix

\section{Approximation for the mass-weighted temperature formula}
\label{sec:appendix_TMW}

The value of the mass-weighted temperature, \tmw, can be calculated by definition as:
\begin{equation}
\begin{split}
    T_{\rm mw} & = \frac{\int_{T_{\rm min}}^{T_{\rm max}} \frac{dM_{\rm dust}}{dT} T dT}{\int_{T_{\rm min}}^{T_{\rm max}} \frac{dM_{\rm dust}}{dT} dT}\\
    & = \frac{1}{M_{\rm dust}} \int_{T_{\rm min}}^{T_{\rm max}} \frac{dM_{\rm dust}}{dT} T dT
\end{split}
\label{eq_appendixTmw_01}
\end{equation}
Using the temperature distribution defined in equation~\ref{eq_Tdist}, the above expression can be rewritten as:
\begin{equation}
    T_{\rm mw} = (1 - {\rm f}_{\rm PDR}) T_{\rm min} + {\rm f}_{\rm PDR} C_{\rm norm} \int_{T_{\rm min}}^{T_{\rm max}} T^{-s+1} dT
\label{eq_appendixTmw_02}
\end{equation}
The second term of the right side, proportional to \fpdr, is the only one which depends on the value of \tmax. Therefore, the \fpdr~=~1 case provides the upper limit to the difference in \tmw~between the truncated and non-truncated temperature distributions. For this reason, we focus on the case \fpdr~=~1 in the following.

Recalling the definition $C_{\rm norm} \equiv(s-1) / \left(T^{1-s}_{\rm min} - T^{1-s}_{\rm max}\right)$ from Section~\ref{sec:methodology:t_dist:t_mw}, we can write:
\begin{equation}
\begin{split}
        T_{\rm mw} & = C_{\rm norm} \int_{T_{\rm min}}^{T_{\rm max}} T^{-s+1} dT \\ 
            & = \frac{s-1}{T^{1-s}_{\rm min} - T^{1-s}_{\rm max}} \int_{T_{\rm min}}^{T_{\rm max}} T^{-s+1} dT \\
            & = \frac{s-1}{s-2} \frac{T^{2-s}_{\rm min} - T^{2-s}_{\rm max}}{T^{1-s}_{\rm min} - T^{1-s}_{\rm max}} \\
            & = \left(1 + \frac{1}{s-2}\right) \frac{(1-x^{s-2})}{(1-x^{s-1})} T_{\rm min}
\end{split}
\label{eq_appendixTmw_03}
\end{equation}
Where we have defined $x \equiv T_{\rm min} / T_{\rm max}$ and we have made use of the fact that $s \neq 0$, $s \neq 1$ in every physically plausible scenario (Section~\ref{sec:methodology:t_dist}), so that we can assume $\int T^{-n} dT = -\frac{1}{n-1}T^{-n+1}$. Since $x < 1$ and $s - 2 > 0$ for our models, we can make use of the relation:
\begin{equation}
    1-x^{s-2} < \frac{(1-x^{s-2})}{(1-x^{s-1})} < 1
\label{eq_appendixTmw_04}
\end{equation}
Combining this with the result of equation~\ref{eq_appendixTmw_03}, we obtain:
\begin{equation}
    \left(1 + \frac{1}{s-2}\right) (1-x^{s-2})\, T_{\rm min} < T_{\rm mw} < \left(1 + \frac{1}{s-2}\right) T_{\rm min}
\end{equation}

In the present paper, it is always true that s~>~6 and \tmin/\tmax~<~1/3, so that we have $x^{s-2} < 1/3^4 = 1/81 \sim 1.2\%$. Therefore, the difference in \tmw\ between truncated and non-truncated temperature distributions is smaller than $\sim 1.2\%$ in the worst-case scenario (\fpdr~=~1), and much smaller in most cases.

\section{Opacity interpolation for carbon data}
\label{sec:appendix_opacity_interpolation}

The opacity $\kappa(\lambda)$ of the materials from \citet{Mennella+98} is measured at five discrete temperatures: 24, 100, 160, 200 and 295 K. Here, we describe the way we interpolated these opacities of the material to cover all intermediate temperature in the range. 

The material opacities in \cite{Mennella+98}, unlike those in \cite{Demyk+17A,Demyk+17B}, are well approximated by a power law: while they do have spectral features in the far-infrared/submillimetre range, they are relatively narrow and weak, and we can expect them to have little effect on our synthetic broad-band photometry. We decided therefore to ignore these features. A power law opacity means that $\log_{10}(\kappa$) is linear in $\log_{10}(\lambda)$. After some experimentation with mathematical models, we decided to approximate the log-opacity as a quadratic function of temperature, as a compromise between precision and computation time. The final form for our interpolation model is therefore: 
\begin{multline}
\label{eq_C_opac}
    \log_{10}(\kappa) = {\rm C}_{0,0} + {\rm C}_{1,0}\, \log_{10}(\lambda) + {\rm C}_{0,1}\, {\rm T} + \\+ {\rm C}_{1,1}\, \log_{10}(\lambda)\, {\rm T} + {\rm C}_{0,2} {\rm T}^2 + {\rm C}_{1,2}\, \log_{10}(\lambda)\, {\rm T}^2
\end{multline} 

\begin{figure}
\begin{center}
\includegraphics[width=.99\columnwidth]{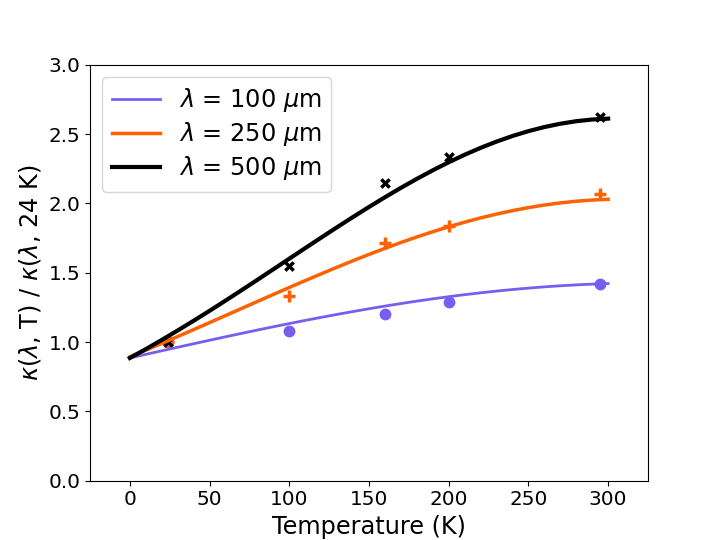}
\caption{
Opacity interpolation for BE carbon, shown for three different wavelengths (100, 250 and 500~$\mu$m). The data points are the experimental measurements at five different temperatures (24, 100, 160, 200 and 295 K), normalized to the value at 24 K for visibility. The curves show the interpolated $\kappa$ as a function of temperature.
}
\label{fig:appendix_opacity_interpolation}
\end{center}
\end{figure}

For the carbon material we use in the present work \citep[BE amorphous carbon;][]{Mennella+98}, and expressing $\lambda$ in $\mu$m, T in K and $\kappa$ in ${\rm cm}^2 \, {g}^{-1}$, the best-fit parameter values are shown in Table~\ref{tab:BE_params}. At the temperatures used for experimental measurement, the model opacity from equation~\ref{eq_C_opac} differs by less than 10\% from the original experimental data for BE carbon, with the exception of a few narrow features. 
For a fixed temperature $T_0$, equation~\ref{eq_C_opac} reduces to a power law with $\beta = - ({\rm C}_{1,0} + {\rm C}_{1,1} T_0 + {\rm C}_{1,2} T_0^2)$. The dependence of opacity on $T$ as expressed by equation~\ref{eq_C_opac} is shown in Fig.~\ref{fig:appendix_opacity_interpolation}, which plots $\kappa(T)$ at three fixed wavelengths ($\lambda = 100$, 250 and 500 $\mu$m) against the experimental data.

\begin{table}
\centering
\caption{
Fit parameters for the opacity of BE carbon according to formula  \ref{eq_C_opac}. Parameters are shown both before and after the opacity correction for aggregation enhancement (see Section~\ref{sec:methodology:opacity}), which affects the values of C$_{0,0}$ and C$_{1,0}$ since the correction is temperature-independent.
}
\begin{tabular}{lll}
\hline
Parameter & Value before  & Value after  \\
          &  correction   &  correction  \\
\hline\hline
C$_{0,0}$ & $\ \ \, 4.86$                & $\ \ \, 4.75$ \\ 
C$_{1,0}$ & $-1.14$               & $-1.29$ \\ 
C$_{0,1}$ & $-3.93 \cdot 10^{-3}$ & ~~(same) \\ 
C$_{1,1}$ & $\ \ \, 6.98 \cdot 10^{-6}$  & ~~(same) \\ 
C$_{0,2}$ & $\ \ \, 2.60 \cdot 10^{-3}$  & ~~(same) \\ 
C$_{1,2}$ & $-4.46 \cdot 10^{-6}$ & ~~(same) \\ 
    \hline
\end{tabular}
\label{tab:BE_params}
\end{table}

Equation~\ref{eq_C_opac} can be extrapolated down to $T$~=~0~K, although values for temperatures below 24~K (the lowest actual temperature from the experimental data) should be used with caution. In Paper~I, we solved this extrapolation issue by assuming -- based on data from different materials -- that carbon opacity would remain fixed below 24~K. In the present work, we relax the assumption to simplify calculations and we limit the temperature range to $T \geq 20$~K, since the change in opacity between 24~K and 20~K is smaller than the $\sim10\%$ uncertainty in opacity itself.

\section{Effect of the temperature distribution slope $s$ on the fit results}
\label{sec:appendix_s_effect}

\begin{figure}
\begin{center}
\includegraphics[width=.99\columnwidth]{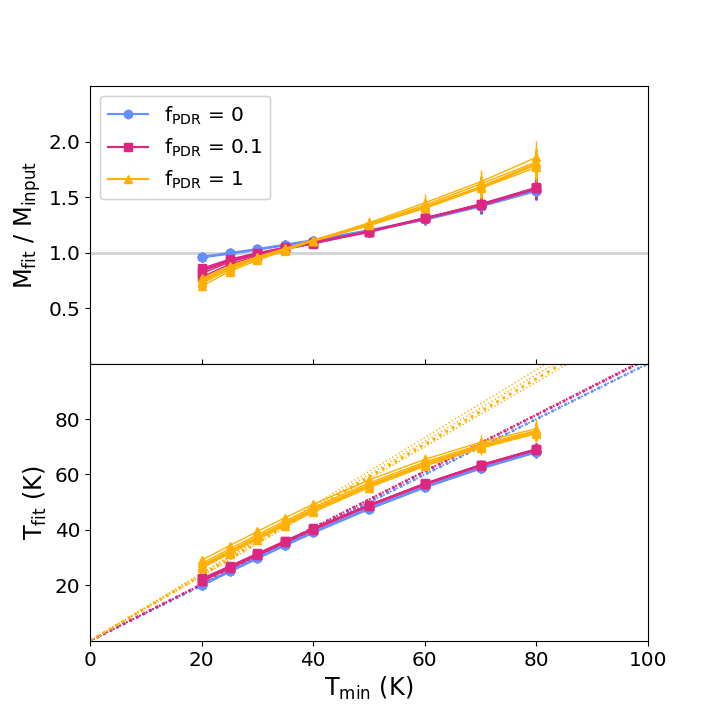}
\caption{
Same as Fig.~\ref{fig:fit_res_fixbeta} (mass and temperature fit results for fixed $\beta$), with additional variations of the index $s$ between 6.5 and 8. In the top panel (\mfit\/\minput), higher curves correspond to higher $s$. In middle panel (\tfit), higher curves correspond to lower $s$. 
}
\label{fig:appendix_s-effects_fixbeta}
\end{center}
\end{figure}

\begin{figure}
\begin{center}
\includegraphics[width=.99\columnwidth]{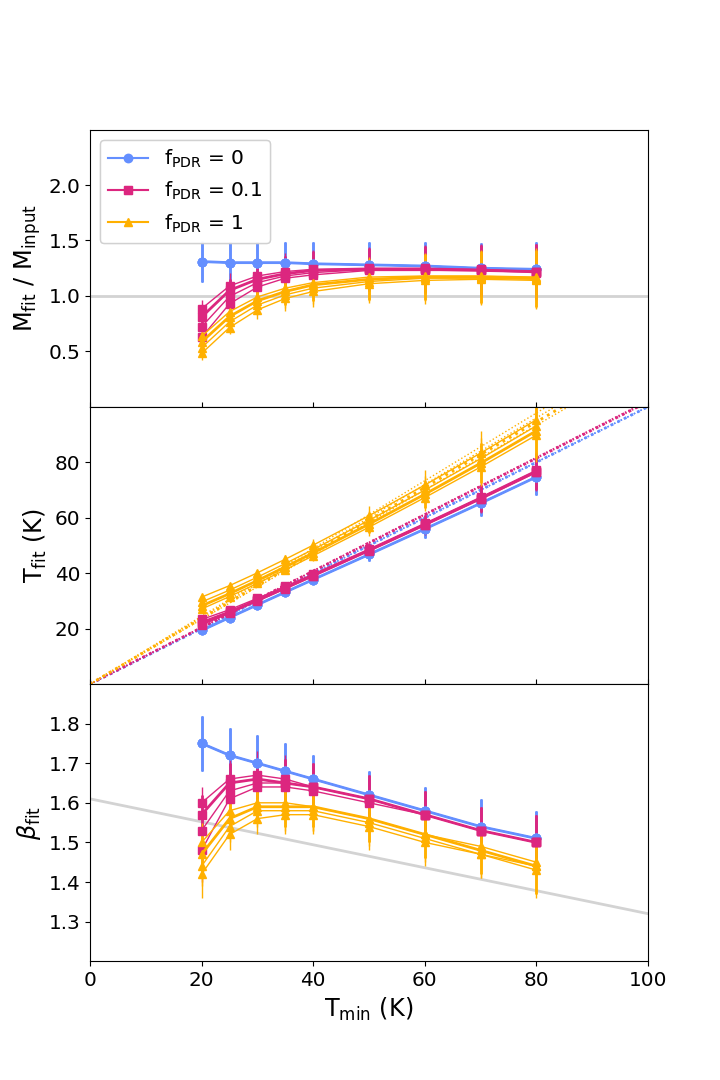}
\caption{
Same as Fig.~\ref{fig:fit_res_fixbeta} (mass, temperature and $\beta$ fit results for free $\beta$), with additional variations of the index $s$ between 6.5 and 8. In the top and bottom panels (\mfit\/\minput\ and \bfit), higher curves correspond to higher $s$. In middle panel (\tfit), higher curves correspond to lower $s$. 
}
\label{fig:appendix_s-effects_freebeta}
\end{center}
\end{figure}

As we discussed in Section~\ref{sec:methodology:t_dist}, under the assumption that dust temperatures are distributed as a power law of index $s$, the expected value for $s$ lies approximately in the $\sim$ 6.5 -- 7.5 range for optically thin media \citep{Kovacs+10}. In the main body of the article we chose to explain only the $s = 7.5$ case for the sake of simplicity. In this section we show that the main thrust of our results does not significantly change if we let the $s$ vary between 6.5 and 8.

We repeated the modified blackbody fits from Section~\ref{sec:results:fixbeta} and \ref{sec:results:free-beta} using different values of $s$. The results are shown in Fig.~\ref{fig:appendix_s-effects_fixbeta} for fixed $\beta$ and Fig.~\ref{fig:appendix_s-effects_freebeta} for free $\beta$, as counterparts of Figs.~\ref{fig:fit_res_fixbeta} and \ref{fig:fit_res_freebeta} from the main article. 
Qualitatively, the effect of increasing (decreasing) $s$ is similar to that of decreasing (increasing) \fpdr: the derived dust mass and $\beta$ are higher (lower) while derived dust temperatures are lower (higher). However, the effect is overall modest. The results of fixed-$\beta$ fits are almost unaffected; for free-$\beta$ fits, the effect is larger but still moderate, and does not qualitatively affect our fit results. The effect of changing $s$ is largest at low \tmin\ and at intermediate star-formation rates (\fpdr~=~0.1).

\bsp	
\label{lastpage}
\end{document}